\documentclass[useAMS,usenatbib]{mn2e}
\usepackage{amssymb,amsmath} 
\usepackage[pdftex]{graphicx}
 \usepackage{graphicx}
 \usepackage{graphics}

\usepackage{epsf}
\usepackage{color}
\usepackage{graphicx}
\usepackage{graphics}
\usepackage{epsfig}
\usepackage{longtable}
\usepackage{multirow}

\usepackage{graphicx, color, setspace, ulem, natbib}

\newcommand{\msun}{\mbox{$M_{\sun}$}}

\newcommand{\numax}{\mbox{$\nu_{\rm max}$}}
\newcommand{\Dnu}{\mbox{$\Delta \nu$}}

\newcommand{\muHz}{\mbox{$\mu$Hz}}
\newcommand{\kep}{\mbox{\textit{Kepler}}}

\title[Asteroseismology of 1523 misclassified red giants]
{Asteroseismology of 1523 misclassified red giants using \kep~data}

\author[Jie.Yu et al.]{Jie Yu$^{1,2}$, Daniel Huber$^{1,2,3}$, Timothy R. Bedding$^{1,2}$, Dennis Stello$^{1,2}$, 	
	Simon J. Murphy$^{1,2}$,\and Maosheng Xiang$^{4}$, Shaolan Bi$^{5}$ and Tanda Li$^{1,2,6}$\\
\\
$^{1}$Sydney Institute for Astronomy (SIfA), School of Physics, University of Sydney, NSW 2006, Australia\\
$^{2}$Stellar Astrophysics Centre, Department of Physics and Astronomy, Aarhus University, Ny Munkegade 120, DK-8000 Aarhus C, Denmark\\
$^{3}$SETI Institute, 189 Bernardo Avenue, Mountain View, CA 94043, USA\\
$^{4}$National Astronomical Observatories, Chinese Academy of Sciences, Beijing 100012, P. R. China\\
$^{5}$Department of Astronomy, Beijing Normal University, Beijing 100875, China \\
$^{6}$Key Laboratory of Solar Activity, National Astronomical Observatories, Chinese Academy of Science, Beijing 100012, China\\}

\begin{document}

\date{Accepted --. Received --; in original form --}


\maketitle

\label{firstpage}

\begin{abstract}
We analysed solar-like oscillations in 1523 \kep~red giants which have previously been 
misclassified as subgiants, with predicted \numax~values (based on the Kepler Input Catalogue) between 280 \muHz~to 700 \muHz. 
We report the discovery of 626 new oscillating red giants in our sample, in addition to 897  
oscillators that were previously characterized by \citet{Hekker11} from one quarter of \kep~data.
Our sample increases the known number of oscillating low-luminosity red giants by 26\%~(up to $\sim$~1900 stars). 
About three quarters of our sample are classified as ascending red-giant-branch stars, 
while the remainder are red-clump stars. A novel scheme was applied
 to determine \Dnu~for 108 stars with \numax~close to the Nyquist frequency 
($240~\muHz<\numax<320\muHz$). Additionally, we identified 47 stars oscillating in the super-Nyquist 
frequency regime, up to 387\muHz, using long-cadence light curves. We show that the misclassifications 
are most likely due to large uncertainties in KIC surface gravities, and do not result from 
the absence of broadband colors or from different physical properties such as reddening, spatial 
distribution, mass or metallicity. The sample will be valuable to study oscillations in low-luminosity 
red giants and to characterize planet candidates around those stars.

\end{abstract}

\section{Introduction}
The advent of space-borne missions such as WIRE \citep{Buzasi00}, MOST \citep{Walker03}, SMEI \citep{Tarrant07}, 
CoRoT \citep{Michel08} and \kep~\citep{Gilliland10} have revolutionized the study of oscillations driven by turbulent near-surface convection (so-called 
solar-like oscillations, \citet{Chaplin13}). Solar-like oscillations exhibit radial and non-radial modes excited over multiple overtones, with amplitudes roughly 
following a Gaussian distribution.
The long, continuous and high-quality light curves from \kep~provide seismic parameters such as the frequency of maximum oscillation power (\numax) and the mean large 
frequency separation (\Dnu). Those can be used in combination with scaling relations to derive stellar properties such as 
masses, radii, densities and even ages, provided complementary spectroscopic and/or photometric information is available  \citep{Stello09, Kallinger10}. Additionally, asteroseismology has allowed new insights into the interior structure, rotation, and magnetic fields of red giants \citep{Bedding11, Beck12, Mosser12a, Fuller15, Stello16a}. 

So far, seismic parameters have been determined for some 500 main-sequence and sub-giant stars 
\citep{Chaplin11, Chaplin14a} and for over 15,000 red giants \citep{Huber10, Huber11, Hekker11, Mosser12b, Stello13}, most of which had previously been identified as red giants 
in the Kepler Input Catalog (KIC, \citet{Brown11}). 
\citet{Huber14} identified 2762 new oscillating red giants which were previously unclassified in the KIC and consolidated stellar characterizations of  
196,468 targets observed by the \kep~Mission. 

In this paper, we are motivated by the fact that some red giants might be misclassified as subgiants in the KIC. 
 While the KIC was successful in its primary goal to distinguish dwarfs from giants, some studies have indicated 
that KIC stellar parameters are significantly biased for subgiants \citep{Molenda11,Verner11a, Thygesen12, Mathur16}. The second motivation stemmed from recent work
by \citet{chaplin14b}, who found that oscillation frequencies located as high as $\sim$ 500 \muHz~can be detected with long-cadence
(29.4 minutes) data, for which the Nyquist frequency is 283.2 $\muHz$. This followed on from work by \citet{Simon13}, who argued that Kepler's 
Nyquist aliases are split into multiplets which allow to discriminate real oscillations from the aliased counterpart. Hence, additional asteroseismic 
red giants may be identified in the super-Nyquist regime.
\section{Sample Selection}
We selected 4758 candidates with KIC effective temperatures less than 6000 K and KIC predicted \numax~ranging 
from near the Nyquist frequency (280 \muHz) to 700 \muHz. The predicted \numax~values were calculated using the scaling relation \citep{Brown91}:
  \begin{equation} 
  {{\nu _{\rm max}}\over {\nu _{\rm max,\odot }}}\simeq {{g/g_{\odot }}\over \sqrt{(T_{\rm eff}/T_{{\rm eff},\odot })}}.
  \label{numaxpre}
  \end{equation}
We used KIC surface gravities \citep{Brown11} and adopted solar values $\nu_{\rm max \odot}$~=~3050~\muHz,  $g_{\odot }$~=~27,487~cm/s$^2$ and $T_{\rm eff \odot}$~=~5777~K. 
We omitted targets with  predicted \numax~larger than 700 \muHz~due to 
the low expected rate of seismic detections, given that those stars are possibly dwarfs or subgiants based on their effective temperatures.
\section{Data Analysis}
\subsection{Full sample}
We have searched all candidates for solar-like oscillations using long-cadence data \citep{Jenkins10} obtained during the full \kep~Mission (Q0 to Q17). 
We used simple aperture photometry (SAP) for our analysis.

Raw light curves were prepared following the procedure described by \citet{garcia11}. The safe-modes have been cut out and instrumental
flux discontinuities were corrected using a linear fit. In order to remove signals due to low-frequency stellar activity and instrumental 
artifacts, we applied a quadratic Savitzky-Golay filter with a length of 10 days. 
We calculated the power spectrum for each target and checked them by eye for the presence of oscillations.
\begin{figure*}
\begin{center}
\includegraphics[width=\textwidth]{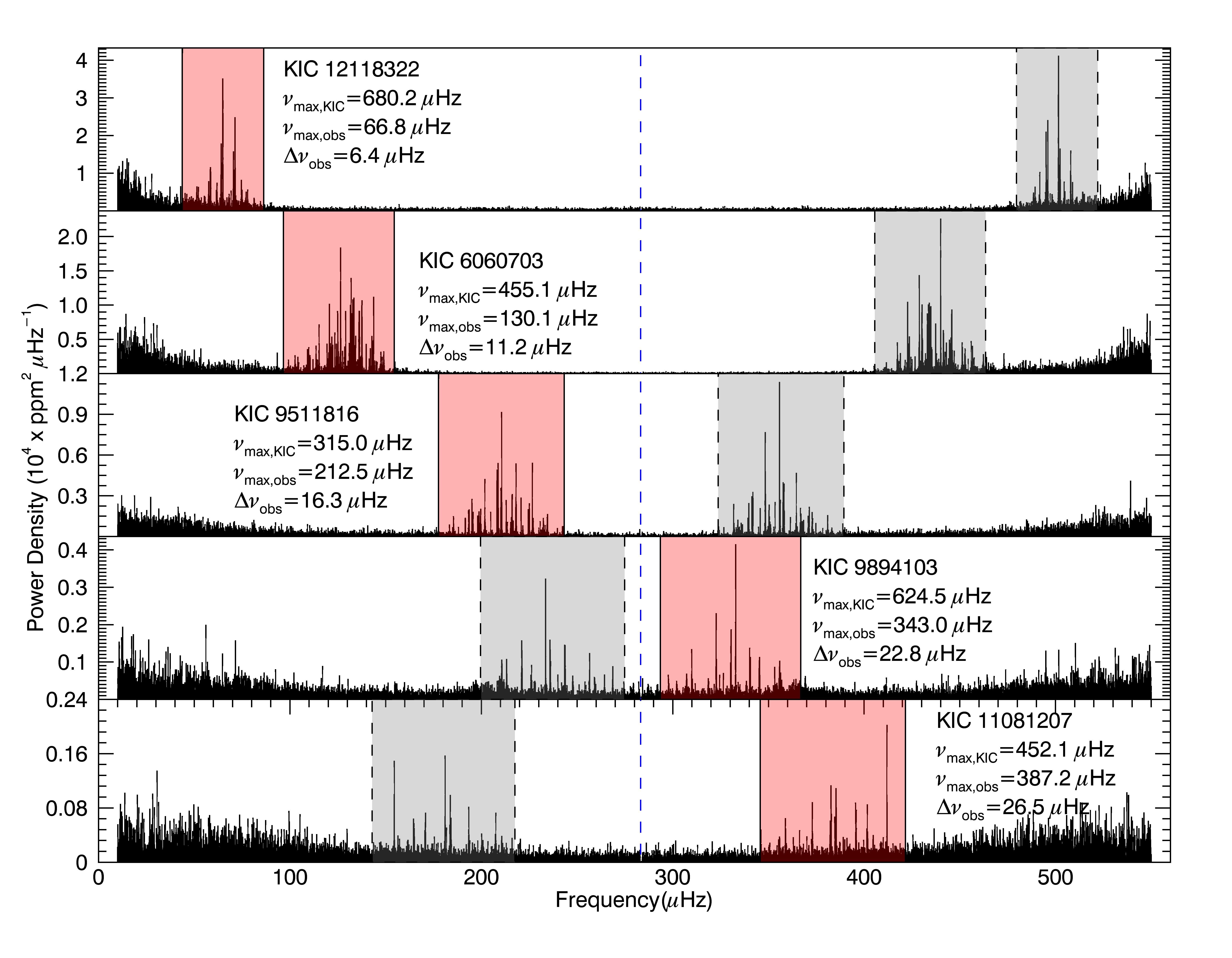}
\caption{Power spectra of five representative red giants. The blue dashed line denotes the Nyquist frequency (283.22 \muHz). The red shaded region between solid lines represents the real oscillation 
power excess, while grey shaded area between dashed lines shows its aliased counterpart. Each star is labeled by KIC number, predicted \numax~(using equation \ref{numaxpre} in combination with 
KIC effective temperature and surface gravity), observed \numax~and \Dnu.}
\label{f5stars}
\end{center}
\end{figure*}
\begin{figure*}
\begin{center}
\includegraphics[scale=0.8]{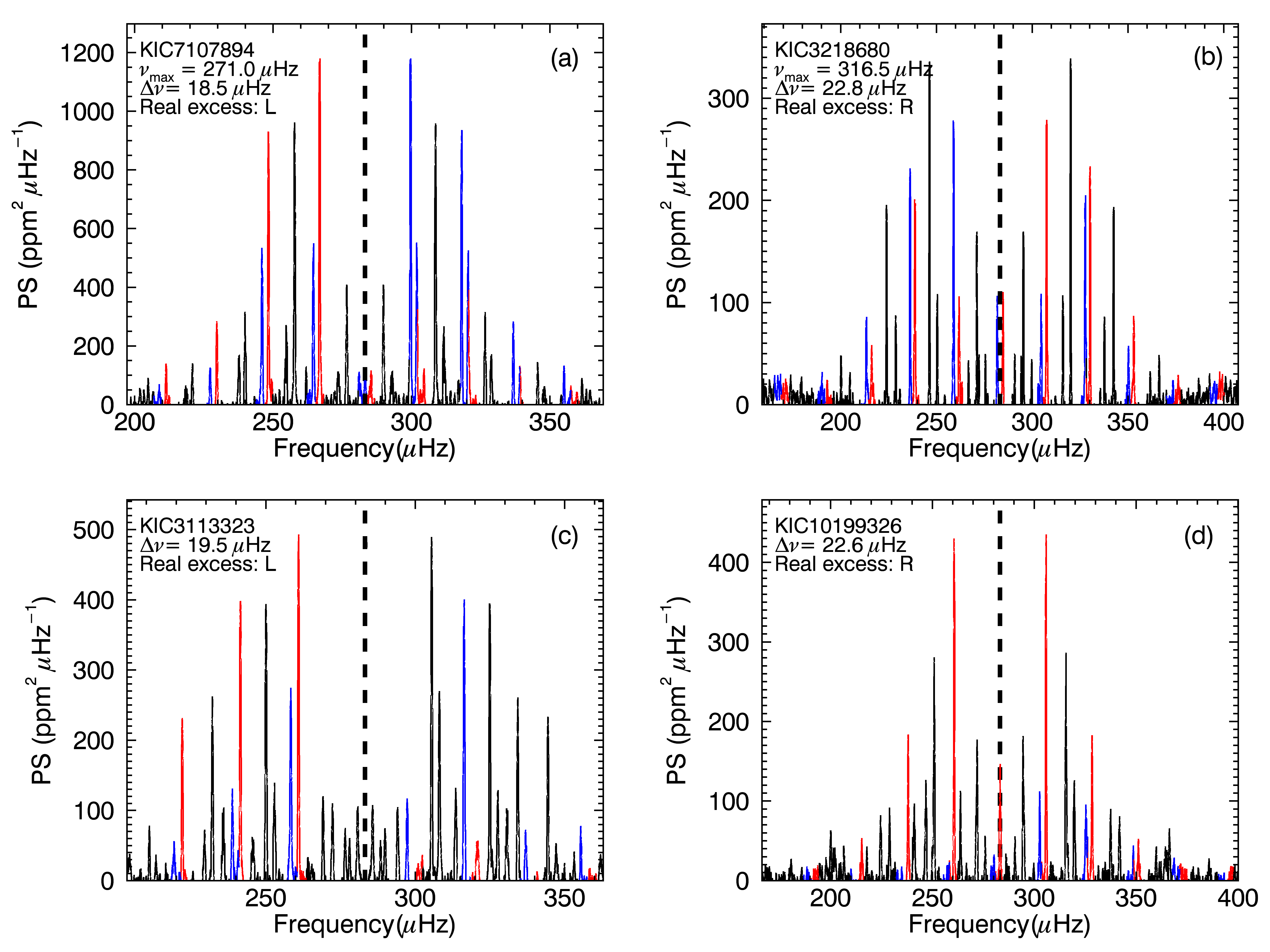}
\caption{Identification of the real power excess for four representative stars oscillating near the Nyquist frequency. The power spectra have been smoothed with 
a boxcar filter with a width of 0.5 \muHz~and the background has been subtracted. Red indicates the regions identified by radial modes while blue shows quadrupole modes. Each target is labeled 
by its KIC ID, observed \numax~(if correctly measured from the pipeline),~\Dnu~and real power excess (L stands for power excess below the Nyquist frequency while R stands for power excess above the Nyquist frequency).}
\label{figplot}
\end{center}
\end{figure*}
Finally, we used the SYD pipeline \citep{Huber09} to extract the global oscillation parameters, including \numax, \Dnu~and amplitude at \numax, 
and calculated the corresponding uncertainties using the same scheme as \citet{Huber11}.

We followed the original method to measure \numax~below the Nyquist frequency. To determine \numax~
above the Nyquist frequency, we refitted the background rather than simply using the reflected background, considering the 
aliased power spectrum as a superposition of the reflected power and real power in the super-Nyquist regime. We tested this method by selecting stars for which both long-cadence and short-cadence data are available, and found good agreement (to within $\sim$~1\%) in the seismic parameters from the two datasets.
                              
Considering that the predicted \numax~values for the targets in our sample are above the Nyquist frequency, it is important to be able to distinguish the real power excess from its aliased counterpart. 
Indeed, one of our aims is to explore the seismic oscillations in the super-Nyquist frequency region using the long-cadence data \citep{Chaplin14a}. 
Therefore, we applied the following approach to identify the real oscillations in
the power spectrum. In its original form, the \citet{Huber09} pipeline calculated the auto-correlation function (ACF) below the Nyquist frequency and determined \Dnu~from 
whichever of the ten highest peaks of the ACF was closest to the predicted \Dnu~using scaling relation. 
In this work, we computed the power spectrum for each target to twice the 
Nyquist frequency, heavily smoothed the ACF of this power spectrum, selected the frequencies of the two highest peaks, $\tau _1$ and $\tau _2$, 
and assigned $\frac{2}{3}(\tau _1 + \tau _2)$ as the predicted $\Delta \nu$. We multiplied by the factor 
$\frac{2}{3}$ because $\tau _1$ and $\tau _2$ correspond to \Dnu~and half \Dnu, respectively. Note that this method does not apply to the oscillating stars with suppressed dipole modes, in which case, 
we just selected the highest peak and assigned its frequency, $\tau$,~to the predicted $\Delta \nu$.
Dipole-mode suppressed oscillators were identified using visual inspection.
The resulting $\Dnu$~is then determined by whichever of the ten highest peaks of 
ACF was closest to the predicted $\Delta \nu$. The scheme was applied to both sides of the Nyquist frequency separately. 
Realizing the fact that the frequencies of maximum seismic power are distinct, appearing 
reflected about the Nyquist frequency, we then distinguished the real from the aliased power excess by comparing which set of \numax~and \Dnu~agrees better with the relation
  \begin{equation} 
  {{\Delta \nu} \simeq \alpha (\nu _{\rm max} /  \mu {\rm Hz})^{\beta}},
  \label{dnupred}
  \end{equation}
where $\alpha=0.268~$\muHz,  $\beta=0.758~$\citep{Huber10}.

Figure \ref{f5stars} shows representative power spectra up to twice the Nyquist 
frequency for five of our detected oscillating giants with high signal-to-noise ratios, including the real (red shaded 
region between solid lines) and aliased (grey shaded region between dashed lines) power excesses. 
The blue dashed line marks the Nyquist frequency. Each target is labeled by the KIC number, predicted \numax~based 
on KIC stellar parameters, and our measured \numax~and \Dnu. Note that the KIC predicted \numax~deviates significantly 
from the observed values. The five stars have monotonically increasing \numax, with KIC 12118322 being the most 
evolved star (\numax =66.8 \muHz), while KIC 9894103 and KIC 11081207 oscillate above the Nyquist frequency. 

We can see that the maximum amplitudes decrease and the widths of the oscillation envelope increase with increasing \numax~\citep{Huber11}.
It is known that for each pair of adjacent $l=2$ (quadrupole) 
and $l=0$ (radial) modes, the latter has the greater power and resides at higher frequency. The asymmetry helps confirm the correct identification. This can be seen clearly 
in KIC 9511816, KIC 9894106 and KIC 11081207 due to their larger \Dnu.  
For the aliased envelopes, the power of the radial mode is lower than the quadrupole mode within each pair, contrary to expectations.

We note that for KIC 12118322, 6060703 and 9511816, the amplitudes of the real excess in Fig. \ref{f5stars} appear to be lower than those of the aliased excess, which is contrary 
to the results given by \citet{Simon13}. We found that the application of a high-pass filter and background removal leads to these reduced amplitudes. However, this difference has negligible influence on the remainder of our analysis.

\subsection{Stars oscillating near the Nyquist frequency}
The method described in Section 3.1 does not work well for stars oscillating close to the Nyquist frequency, 
given parameters uncertainties and, in particular, the difficulty in determinating \numax. Hence, we applied a different method for those stars.

It is known that high-order acoustic modes follow the asymptotic relation, $\nu \approx \Dnu (n+l/2+\epsilon)$, where $\epsilon$ 
may be evaluated from \Dnu~based on the relation: $\epsilon = 0.634+0.63 \log(\Dnu)$ \citep{Corsaro12}. The large separation \Dnu~returned from 
the pipeline can therefore be used to predict $\epsilon$ since it is measured independently of \numax~(Sect. 3.1). 

To select the power spectrum regions dominated by $l=0$~and $l=2$, we 
employed the relation: $-0.22<(v/\Dnu~-~\epsilon)\mod 1<-0.06$ for $l=2$ and $-0.06<(v/\Dnu~-~\epsilon)\mod1<0.10$ for $l=0$~\citep{Stello16a, Stello16b}. 
We selected up to 4 pairs of radial and quadrupole modes which 
were closest to the highest peak at both sides of the Nyquist frequency. We did this instead of using \numax~from the 
pipeline since \numax~is less accurate for 
stars oscillating near the Nyquist frequency. 
Both the real and aliased excesses were matched against a fine structure template constructed via the 
$\epsilon-\Dnu$ relation and the one with the higher agreement with the template was selected as the real excess.
However, this scheme does not work well if the real mode pattern falls symmetrically around the Nyquist frequency (i.e. either if an acoustic dipole resonance mode 
or a pair of $l=0,~2$ modes fall right at the Nyquist frequency). In such case,  
we compared the power of radial modes with that of quadrupole modes on either side of the Nyquist frequency. We identify the real excess as that where the modes 
identified as quadrupoles are of lower power than their neighbouring radial modes. 
In practice, we performed this only if $|p_1-p_2|/p_1<20\%$, where $p_1$ and $p_2$ represent the total power dominated by radial and
quadrupole modes below and above the Nyquist frequency, respectively. We selected a slightly conservative threshold (20\%), given that the relations for predicting the 
$l=0$ and $l=2$ dominated frequency regions are not exact.

Figure \ref{figplot} displays four representative stars with \numax~close to the Nyquist frequency. The power spectra have been smoothed to 
a resolution of 0.5 \muHz. Red indicates the regions identified as radial modes from the $\epsilon$-\Dnu~relation while blue shows quadrupole modes. The real power excess 
determined by the scheme agrees well with the result from the pipeline for a star oscillating below the Nyquist 
frequency, as shown in Fig 2a, and a star oscillating in the super-Nyquist region, as shown in Fig 2b, 
and confirms the seismic parameters, in particular \Dnu. Conversely, \numax~for 
the star in panel (c) was overestimated (\numax~returned from the pipeline was 311.6 \muHz) while in panel (d) it was underestimated (\numax~returned from the 
pipeline was 256.8 \muHz). The accuracy of \Dnu~from the pipeline is confirmed after realizing the excellent agreement between the identified and real modes. 
We applied the scheme to 108 stars with \numax~from 240 \muHz~to 320 \muHz~and confirmed the accuracy of \Dnu~by eye.
\section{Results}\label{results}
\begin{figure}
\begin{center}
\includegraphics[scale=0.55]{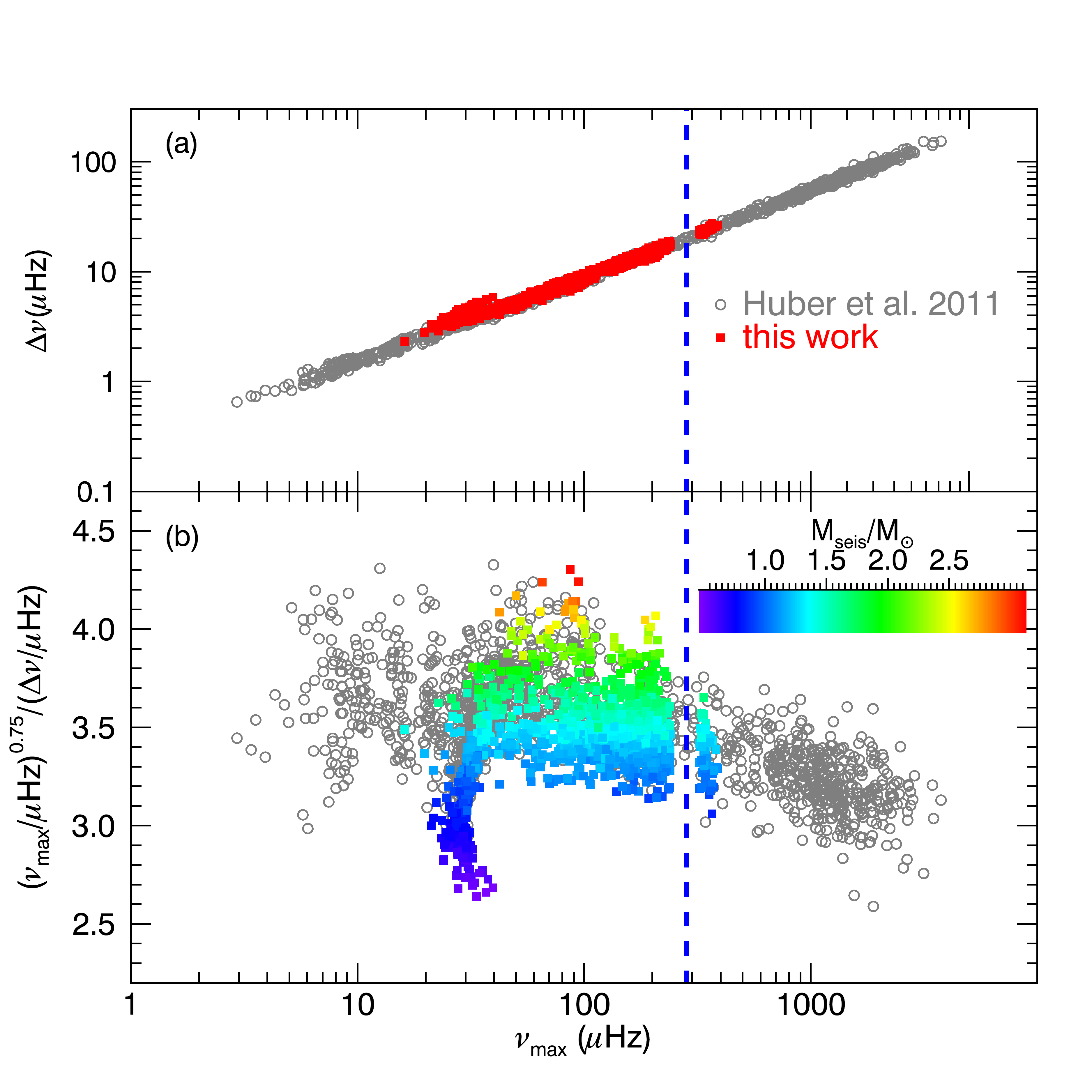}
\caption{(a): \Dnu~vs~\numax~for the entire sample (red squares) and those from \citet{Huber11} (grey circles). 
(b): $\numax ^{0.75}/\Dnu$~vs~$\numax$. The blue dashed line denotes the Nyquist frequency (283.22\muHz). Asteroseimic masses are color-coded, which were derived 
by combining \numax, \Dnu~and KIC effective temperature. We can see clearly the diversity of masses dominates the scatter in the vertical 
direction. Note that we excluded the stars with measured \numax~in the range of 240 \muHz~and 320 \muHz~(see text).}
\label{fnumaxdnu}
\end{center}
\end{figure}

\begin{figure}
\begin{center}
\includegraphics[scale=0.52]{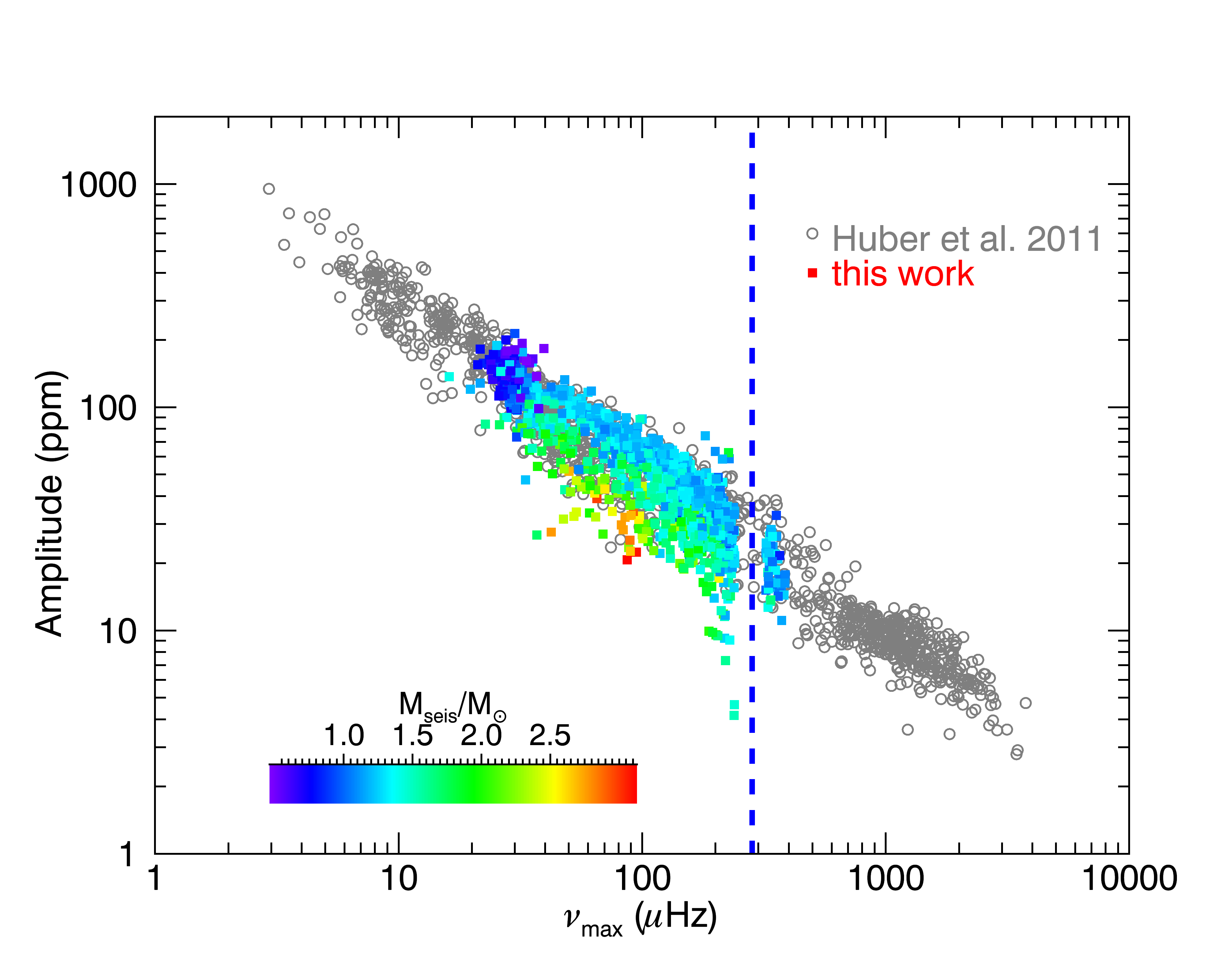}
\caption{Oscillation amplitude at \numax~against~\numax~for the whole sample (color-coded by the seismic mass). Targets from \citet{Huber11} are plotted as grey circles.
The blue dashed line denotes the Nyquist frequency (283.22\muHz). Targets oscillating near the Nyquist frequency have underestimated 
amplitudes (see text).}
\label{famplitude}
\end{center}
\end{figure}
In this section we present our main results for the entire sample and compare our detections to the sample of \citet{Huber11}.  
Our sample consists of 1523 stars, all of which have oscillations that are clearly visible in the power spectra. After excluding outliers and some targets which have been previously analyzed, we report the discovery of 626 new oscillating red giants. The majority of known red giants were investigated by \citet{Hekker11}, who analysed solar-like oscillations in those red giants by using only one quarter of \kep~data. The application of data from full \kep~Mission in our work enables us to measure seismic parameters more precisely. We list global seismic parameters, \numax, \Dnu~and the amplitude per radial mode \citep{Kjeldsen08, Huber10} in Table~\ref{tab:parameter}. Note that \numax~values for stars between 240 \muHz~and 320 \muHz~were omitted because of the difficulty of precisely measuring \numax~near the Nyquist frequency.
\subsection{Global asteroseismic parameters: \numax, \Dnu~and Amplitude}
Figure \ref{fnumaxdnu} illustrates the global asteroseismic parameters, \numax~and \Dnu~for the whole sample (red squares) and compared with the sample 
of \citet{Huber11} (grey circles). \numax~and~\Dnu~follow the well-known power law 
relation as given in equation \ref{dnupred}. The fitted coefficients, $\alpha = 0.262\pm0.001$ and $\beta = 0.770\pm0.001$, are consistent with \citet{Huber11} and almost identical to the original fit by \citet{Stello09}. Uncertainties were calculated using a Monte Carlo simulation and are smaller than in the literature due to the longer datasets used in our work.

To reveal the additional information encoded in \numax~and \Dnu~, we took a ratio of \numax~and \Dnu~which is solely dependent on 
mass and temperature \citep{Huber11}:
\begin{equation}
{{(\numax / \muHz)^{0.75} \over {\Dnu /\muHz}} \propto \left({M \over {M _{\odot}}}\right)^{0.25} \left({T _{\rm{eff}} \over {T _{\rm{eff}, \odot}}}\right)^{-0.375}}.    
\label{numaxdnu}
\end{equation}
The result is shown in Fig \ref{fnumaxdnu}b, where asteroseismic masses, which were calculated from \numax, \Dnu~and the KIC effective temperature, are color-coded. 
We can see clearly that the diversity of masses of red giants produces the scatter in the vertical direction. 
We have checked that the effect of effective temperature on the distribution is minor, 
reflecting the fact that red giants, in particular in our sample, have similar effective temperatures. 
The agreement between our sample and those from \citet{Huber11} is good. Note that the red-clump stars, which are in the helium-core burning phase, produce increased number of stars seen in both panels at \numax~around 20-50\muHz~because of the   relatively longer evolutionary lifetime.

\citet{Huber11} also tested the scaling relation for oscillation amplitudes from the main sequence to red giants. They made a comparison between the observed and 
predicted amplitudes which were based on the scaling relation:

\begin{equation}
{A_{Kp} \propto {L^s \over {M^t~T_{\rm{eff}}^{r-1}~C_K(T_{\rm{eff}})}}},
\label{amplitudes}
\end{equation}
where $L$ is the luminosity, $M$ is the mass, $T_{\rm{eff}}$ is the effective temperature, $s=0.838\pm0.002$~and~$t=1.32\pm0.02$~are the fitted coefficients, $r$ 
is fixed at 2 \citep{Huber11}, and $C_K(T_{\rm{eff}})$~is the bolometric correction factor \citep{Ballot11},

\begin{equation}
{C_K(T_{\rm{eff}}) = \left({T_{\rm{eff}} \over {5934K}}\right)^{0.8}}.
\label{CK}
\end{equation}
\citet{Huber11} argued that the spread of amplitudes, which is significantly larger than the uncertainties at a given \numax, is partially 
due to the dispersion of stellar masses and that lower-mass stars show larger amplitudes at a given \numax. In Figure \ref{famplitude} we do see the amplitude dispersion and the 
mass-amplitude relation, both of which have good agreement with those from \citet{Huber11}. We have checked 
the outliers at $\numax \sim 37.0~\muHz$~and found the corresponding amplitude is reliable, indicating that 
the low amplitude may arise from an unresolved companion that dilutes of the oscillation amplitude. 
We can also see a few stars which have relatively lower amplitudes near the Nyquist frequency. Manual inspection showed that this is 
due to the difficulty of determining the photon noise level when the frequency of maximum oscillation power is close to the Nyquist frequency.

\begin{table}
\begin{small}
\begin{center}
\caption{Global oscillation parameters: \numax,~\Dnu~and amplitude.}
\vspace{0.1cm}
\begin{tabular}{l c c c c c c}        
\hline         
         KIC ID     &       \numax~(\muHz) 	  &        \Dnu~(\muHz)     &  Amplitude~(ppm)  \\
\hline 
        2300227     &       61.1(0.4)     &      6.29(0.02) &   93.5(4.3) \\
	4173334     &      125.5(1.7)     &     10.04(0.04) &	24.6(1.7) \\
	5371482     &      169.9(3.5)     &     12.32(0.06) &   23.2(3.2) \\
	5374099     &      133.2(0.9)     &     10.75(0.03) &   40.7(2.2) \\
	6060703     &      130.1(0.6)     &     11.23(0.02) &   51.2(1.4) \\
	8733649     &      148.3(3.4)     &     11.53(0.03) &   19.0(1.3) \\
	9579357     &      133.2(1.0)     &     10.47(0.03) &   26.3(1.6) \\
	9511816     &      212.5(1.2)     &     16.32(0.31) &   23.5(3.6) \\
        9761128     &      361.8(8.7)     &     26.11(0.05) &   14.2(4.7) \\
	9894103     &      343.0(4.2)     &     22.84(0.06) &   26.3(3.5) \\
       11147387     &        ****         &     16.28(0.04) &    ****     \\
\hline
\end{tabular} 
\label{tab:parameter} 
\flushleft Note. Asterisks indicates that the corresponding values are not available due to the \numax~being close to the Nyquist 
frequency (between 240 \muHz~and 320 \muHz). Values in brackets represent uncertainties.
\flushleft (This table is available in its entirety in a machine-readable form in the 
online journal. A portion is shown here for guidance regarding its form and content.)
\end{center}
\end{small}
\end{table}
\subsection{Evolutionary States} \label{population}
It is well known that red giants in the hydrogen-shell-burning phase (RGB) and giants in the helium-core burning phase, 
which are designated as red-clump stars (RC1) and secondary red-clump stars (RC2), overlap in the Hertzsprung-Russell diagram (H-R diagram), making them difficult
to discern using classical observing techniques.
\citet{Bedding11} and \citet{Mosser2012c} proposed that period spacings of mixed dipole modes can be used 
 to disentangle the RGB (period spacings of $\sim$~50 seconds) from RC1 and RC2 (period spacings between 100 to 300 seconds). \citet{Stello13} extended this investigation to classify
 stellar populations amongst 13,000 red giants. Recently, \citet{Mosser15} and \citet{Vrard16} have utilized new methods to measure rotational splittings and period spacings in an automatic way.
\citet{Takeda08} and \citet{Takeda15} adopted an alternative approach by distinguishing the evolutionary states
 in a surface gravity against mass diagram. Since measuring period spacings is beyond the scope of this paper, we
 apply the log~$g$-mass method to explore evolutionary states.
\begin{figure*}  
\begin{center}
\includegraphics[scale=0.65]{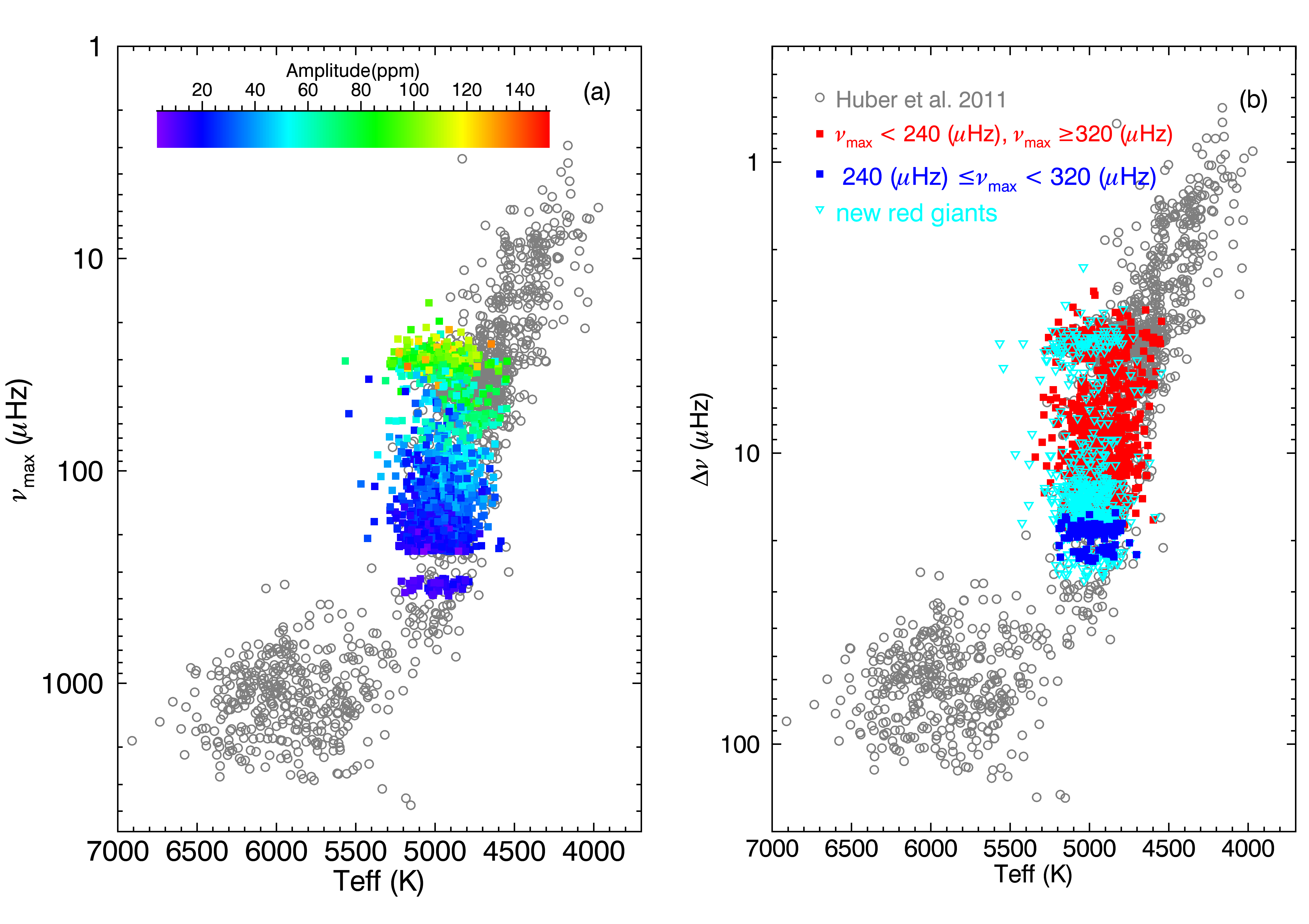}
\caption{(a): \numax~vs $T_{\rm{eff}}$ diagram. The abscissa is the KIC effective temperature. Targets from the entire sample are color coded by their amplitudes. 
Stars from \citet{Huber11} are shown as grey circles for comparison. The gap is due to the absence of \numax~measurements in the range 
240 \muHz~to 320 \muHz. (b): \Dnu~vs $T_{\rm{eff}}$ diagram. Newly discovered oscillating red giants are shown as cyan open triangles. 
Stars with \numax~near the Nyquist frequency are denoted as blue squares.}
\label{fhrdiagram}
\end{center}
\end{figure*}

\begin{figure}
\begin{center}
\includegraphics[scale=0.53]{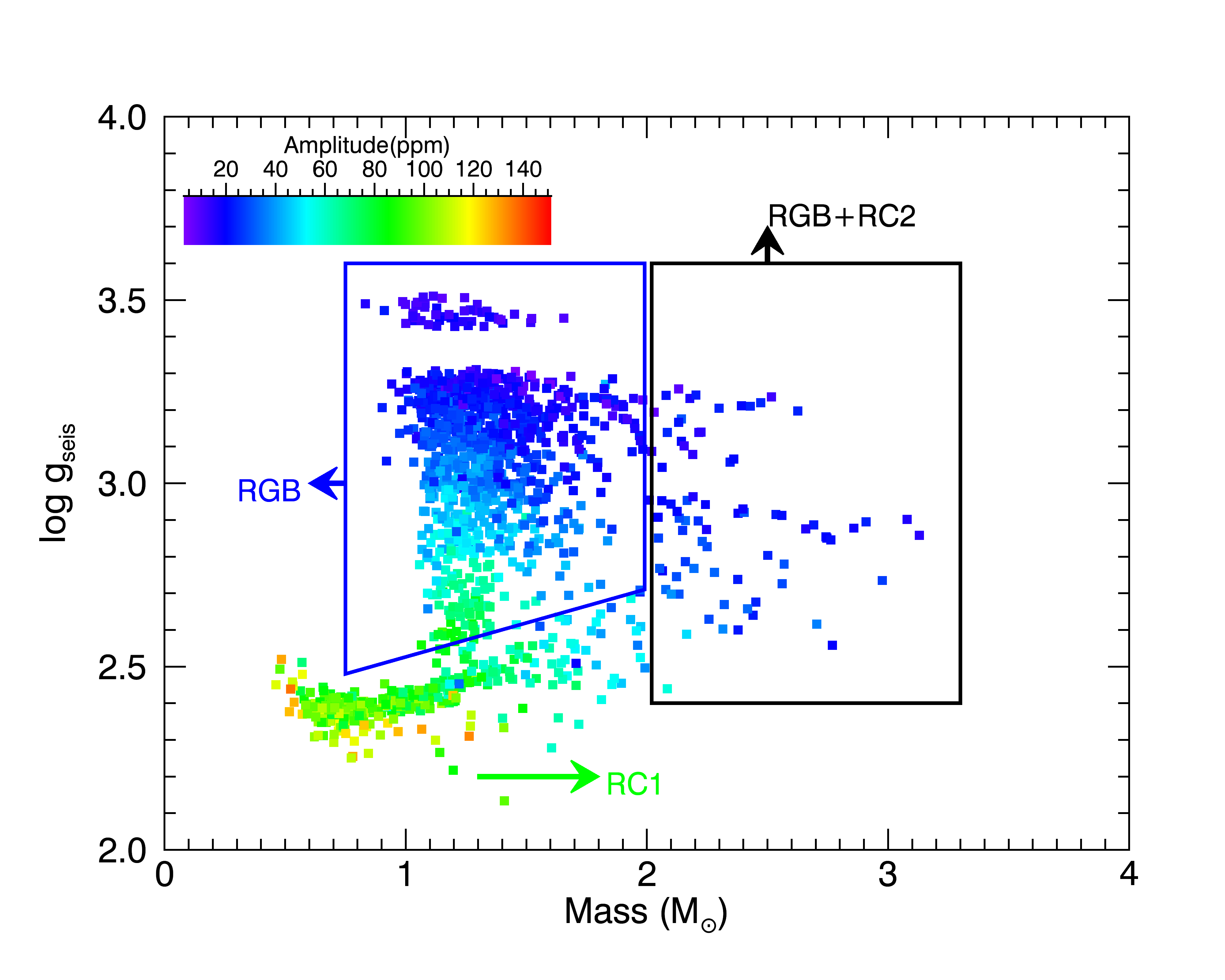}
\caption{Asteroseismic surface gravities versus masses. Targets are color coded by the oscillation amplitude. Boxes label approximate evolutionary stages for RGB and RC2. Stars outside the boxes 
are RC1.}
\label{fseishrdiagram}
\end{center}
\end{figure}

Firstly, as seen in Figure \ref{fnumaxdnu} and Figure \ref{famplitude}, targets in our sample are mainly red giants, which generally have \numax
~less than 300 \muHz. Figure \ref{fhrdiagram}a shows the asteroseismic H-R diagram with each target from our sample color-coded by oscillation amplitude. 
Compared with the sample from \citet{Huber11} our sample lacks more evolved red giants with \numax~less than 20 \muHz, which 
results from our low frequency threshold of 10 \muHz~and from the fact that the KIC classifications are more accurate for cool stars. Again, 
the gap seen in panel (a) is due to the absence of \numax~measurements in the range 240 \muHz~to 320 \muHz. This gap is filled substantially if instead of \numax, we plot \Dnu~as shown in panel (b), where 
newly discovered oscillating red giants are indicated by cyan open triangles.

To further classify different populations in our sample, we plotted the asteroseismic surface gravities versus masses in Figure \ref{fseishrdiagram} with approximate evolutionary 
stages. RGB stars account for $\approx 71\%$ of sample while RC1 stars take up $\approx$ 25$\%$ of sample. The remaining $\approx$ $4.0\%$ of stars have seismic masses larger than 2.0\msun~and can either be classified as 
RGB or RC2 (statistically more likely to be RC2 stars). The higher amplitude stars with surface gravities from 2.5 dex to 2.8 dex and seismic masses from 1.1\msun~to 1.4\msun~(green points) 
are likely higher-luminosity RGB stars or asymptotic giant stars. RC1 stars are also well classified with a bar 
shape at the left-bottom corner, which show the highest amplitudes as color-coded by the green and red squares. 
The RGB stars mostly consist of low-luminosity red giants which have relative low oscillation amplitude as denoted by the blue points. 

Compared to \citet{Huber11} and \citet{Stello13} samples, our sample increases the number of low-luminosity ($\numax>100~\muHz$) red giants by $\sim$~26\% (up to $\sim$~1900 stars).
\begin{figure}
\begin{center}
\includegraphics[width=\columnwidth]{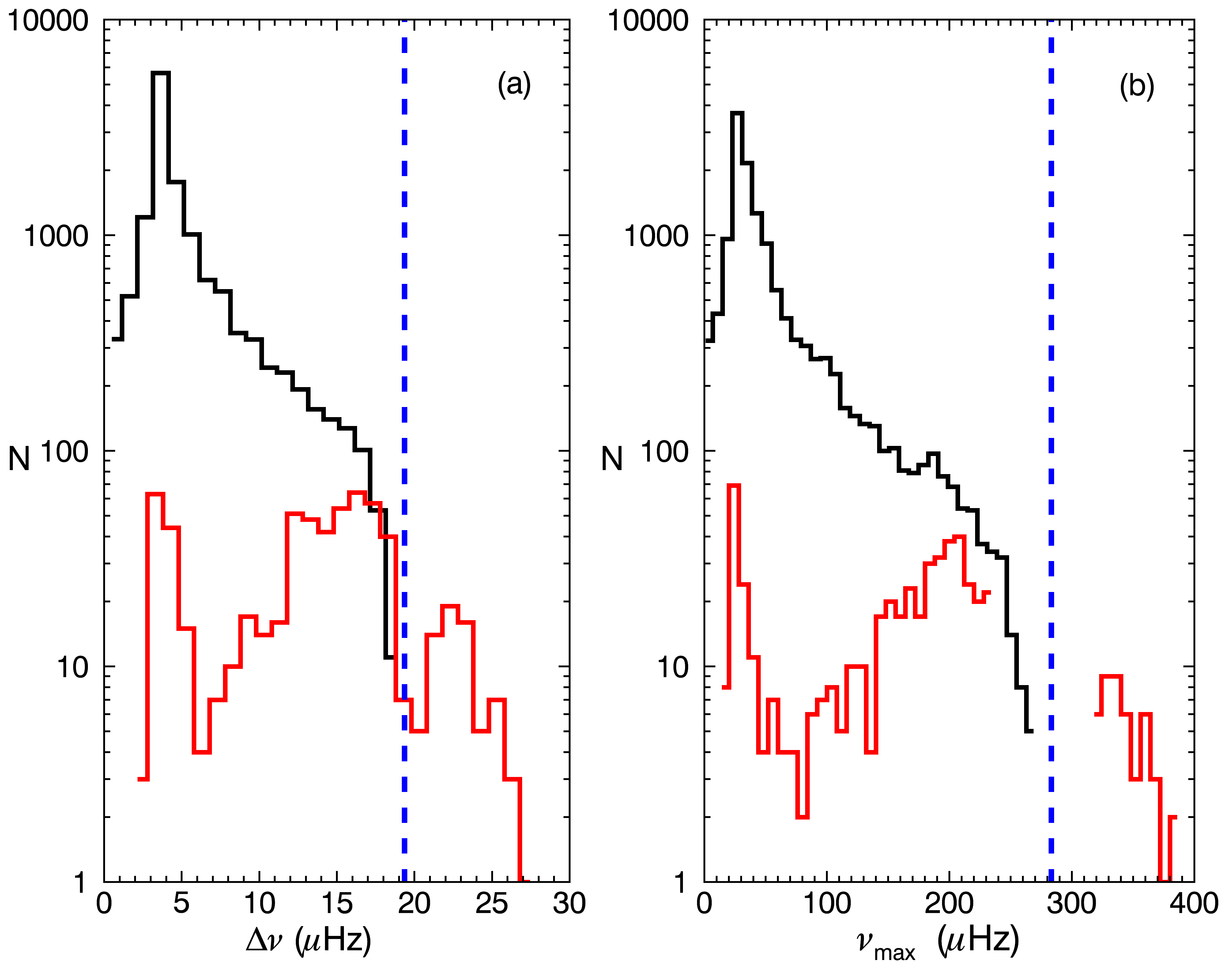}
\caption{Histograms of \Dnu~and \numax. Red curve represents our sample while black curve indicates \citet{Huber11} and \citet{Stello13} samples. The blue dash lines mark the Nyquist frequency in panel (b) and its corresponding \Dnu~calculated from scaling relation in panel (a). The short-cadence sample by \citet{Huber11} is not shown here.}
\label{histdnunumax}
\end{center}
\end{figure}
Figure \ref{histdnunumax} illustrates the histograms of \numax~and \Dnu~for our sample (red curve) as well as the combined \citet{Huber11} and \citet{Stello13} samples (black curve). Panel (a) shows all the targets in our sample, including stars oscillating near the Nyquist frequency, while panel (b) does not display those 
stars. We can see from panel (b) that our sample has a large number of low-luminosity red giants. It also includes 47 stars oscillating beyond the Nyquist frequency, up to 387\muHz.
 
Low-luminosity red giants are characterized by broad power excess and rich modes 
due to the presence of mixed modes and rotational splittings. 
An expanded sample of low-luminosity red giants is therefore very valuable to asteroseismically understand their global properties and internal structures. 
We find 16 stars in our sample are planet-candidate host stars which harbour 17 exoplanet candidates listed in the NASA Exoplanet 
Archive\footnote{http://exoplanetarchive.ipac.caltech.edu/index.html}. Those exoplanet candidates will be investigated in a follow-up paper.
\section{Discussion}
\begin{figure}
\begin{center}
\includegraphics[scale=0.53]{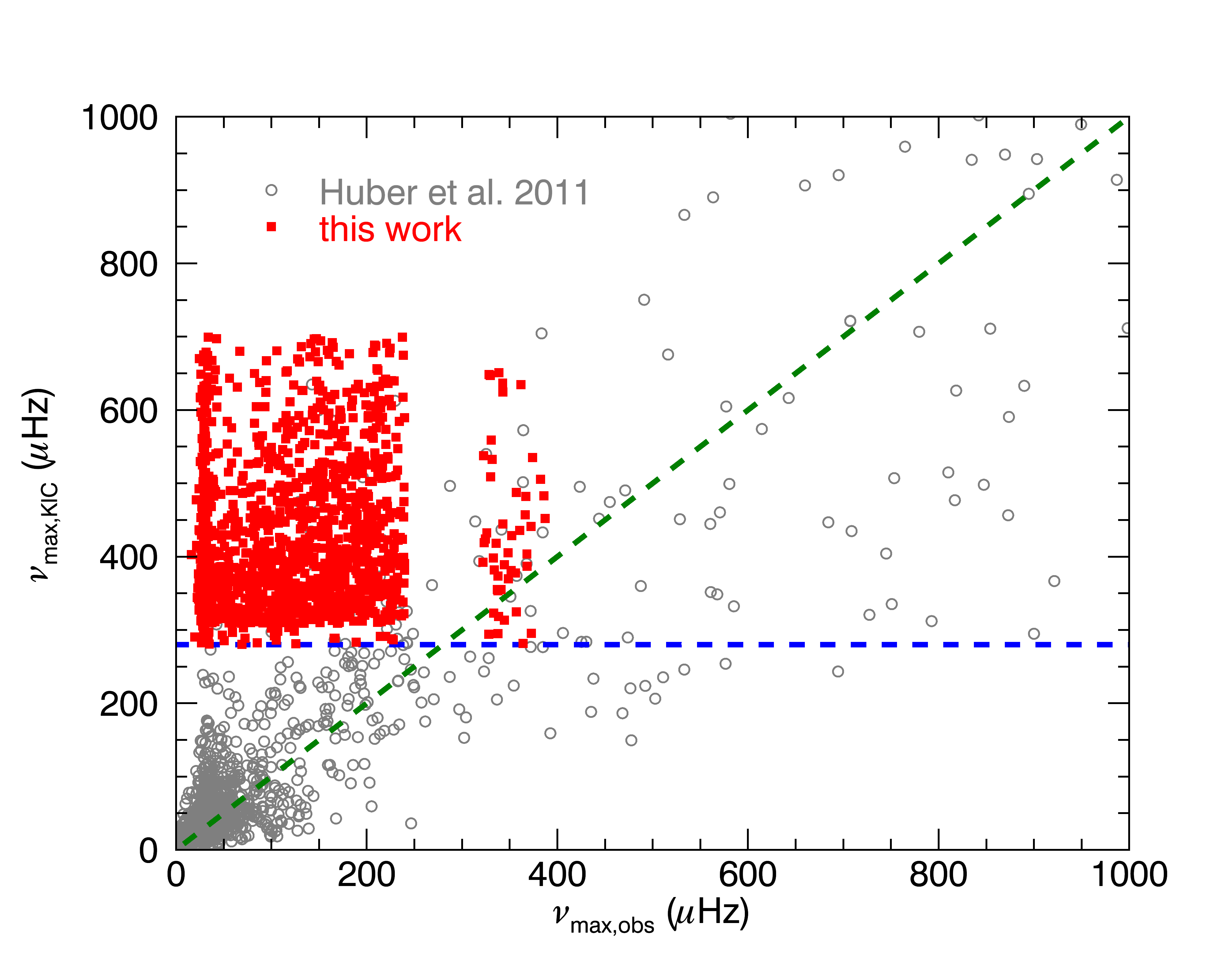}
\caption{KIC-predicted versus observed \numax~for the entire sample (red squares) and those of \citet{Huber11} (grey circles). KIC predicted \numax~values were derived from equation \ref{numaxpre} with 
KIC temperatures and surface gravities. The green dashed line denotes the 1:1 relation, while the blue dashed line represents the lower limit of the predicted \numax~for the sample selection.}
\label{numaxkic}
\end{center}
\end{figure}
As shown in section \ref{results}, our sample is mostly comprised of low-luminosity red giants and red-clump stars, suggesting that the predicted \numax~values~based on the KIC 
are significantly biased compared with the observed \numax. Figure \ref{numaxkic} shows that the KIC and observed \numax~values for 
our sample strongly deviate from the one-to-one relation, while the targets in \citet{Huber11} show better agreement.
In order to understand the cause of the offset in \numax~shown in Figure \ref{numaxkic}, we used spectroscopic data 
to determine whether this offset originates from some particular physical properties in our sample or from random errors. 
\subsection{Comparison with LAMOST}
\label{cmplmst}
\begin{figure}
\begin{center}
\includegraphics[width=\columnwidth]{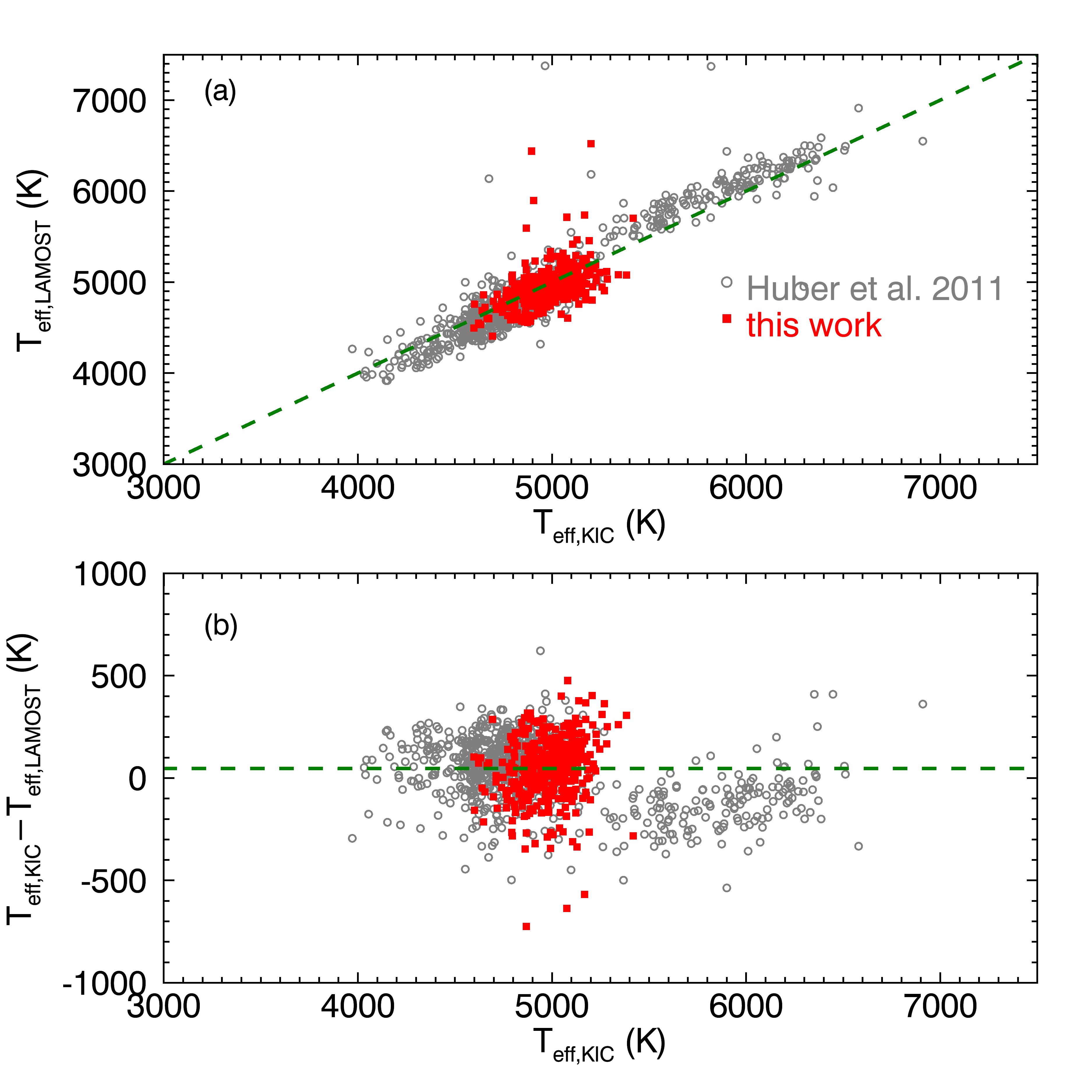}
\caption{Comparison of effective temperature from KIC and LAMOST for our sample (red squares) and those of \citet{Huber11} (grey circles). The green dashed line shows the 1:1 relation in panel (a)  
and the mean difference in the sense of KIC minus LAMOST in panel (b) for our sample.}
\label{teffcmp}
\end{center}
\end{figure}

\begin{figure}
\begin{center}
\includegraphics[width=\columnwidth]{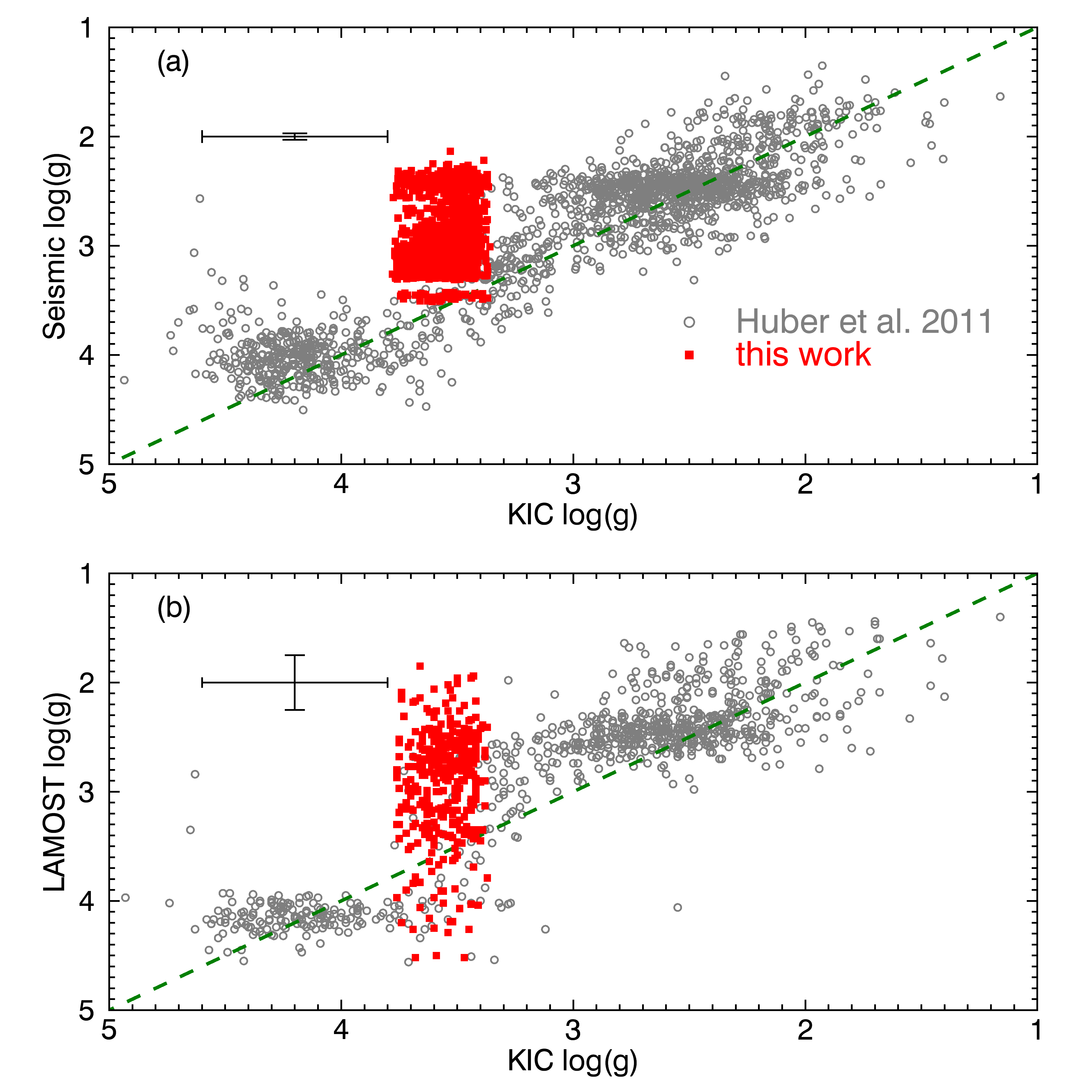}
\caption{(a) Asteroseismic versus KIC surface gravities for the entire sample(red squares) and those of \citet{Huber11} (grey circles). 
Asteroseismic surface gravities are derived using equation \ref{numaxpre} in combination with observed \numax~and KIC effective temperature. (b): Same with panel (a) but for LAMOST. Typical uncertainties 
are shown for each surface-gravity scales in both panels}
\label{seislogg}
\end{center}
\end{figure}

The LAMOST (Large Sky Area Multi-Object Fiber Spectroscopic Telescope, \citet{Zhao12}) survey 
collected low resolution ($R \approx 1800$) optical spectra (3800-9000$\AA$) for objects in the \kep~fields \citep{Cui12,Luo15,De2015}. 
Stellar atmospheric parameters ($\rm{T_{eff}}$, log~$g$, [Fe/H]) of the LAMOST-Kepler spectra collected before September 2014 were derived
with LSP3 pipeline \citep{Xiang15}. The LSP3 pipeline uses template
matching with the MILES stellar spectra library, and
estimates measurement uncertainties of stellar atmospheric parameters for individual
stars based on their S/N and location in the stellar parameter space. Detailed tests have shown that LSP3
provides parameters with an overall accuracy of about 150 K ($\rm{T_{eff}}$), 0.25 dex (log~$g$) and 0.15 dex ([Fe/H]), 
given a spectral signal-to-noise ratio (S/N) higher than 10 \citep{Xiang15}. For F/G
type stars with high spectral S/N, uncertainties can be even smaller than 100 K ($\rm{T_{eff}}$), 0.2 dex (log~$g$) and 0.1 dex ([Fe/H]).

There are 368 common stars between our sample and the LAMOST-Kepler catalog \citep{Xiang15}. We first compared KIC with LAMOST effective temperatures as shown 
in figure \ref{teffcmp}. Panel (a) indicates that there is a good agreement between the two temperature scales, with the KIC temperatures 
being slightly hotter. After removing eight outliers using $4\sigma$ clipping, 
we obtained the mean and scatter of the residuals in our sample (in the sense of KIC minus LAMOST) as $+47K\pm154K$. Inspection of Figure \ref{teffcmp} also reveals a similar offset and spread arising for the red giants from the \citet{Huber11} sample. For main-sequence stars, spectroscopic temperatures from LAMOST are higher, consistent with previous comparisons \citep{Huber14, Pinsonneault12}.

In addition to the effective temperature, surface gravity is the other source of uncertainty for predicting \numax. Asteroseismology provides accurate log~$g$ measurements and hence has been used as input to lift the degeneracy between spectroscopic temperature, surface gravity and metallicity \citep{Bruntt12,Huber13,Pinsonneault14,Liu15}. 

\begin{figure}
\begin{center}
\includegraphics[width=\columnwidth]{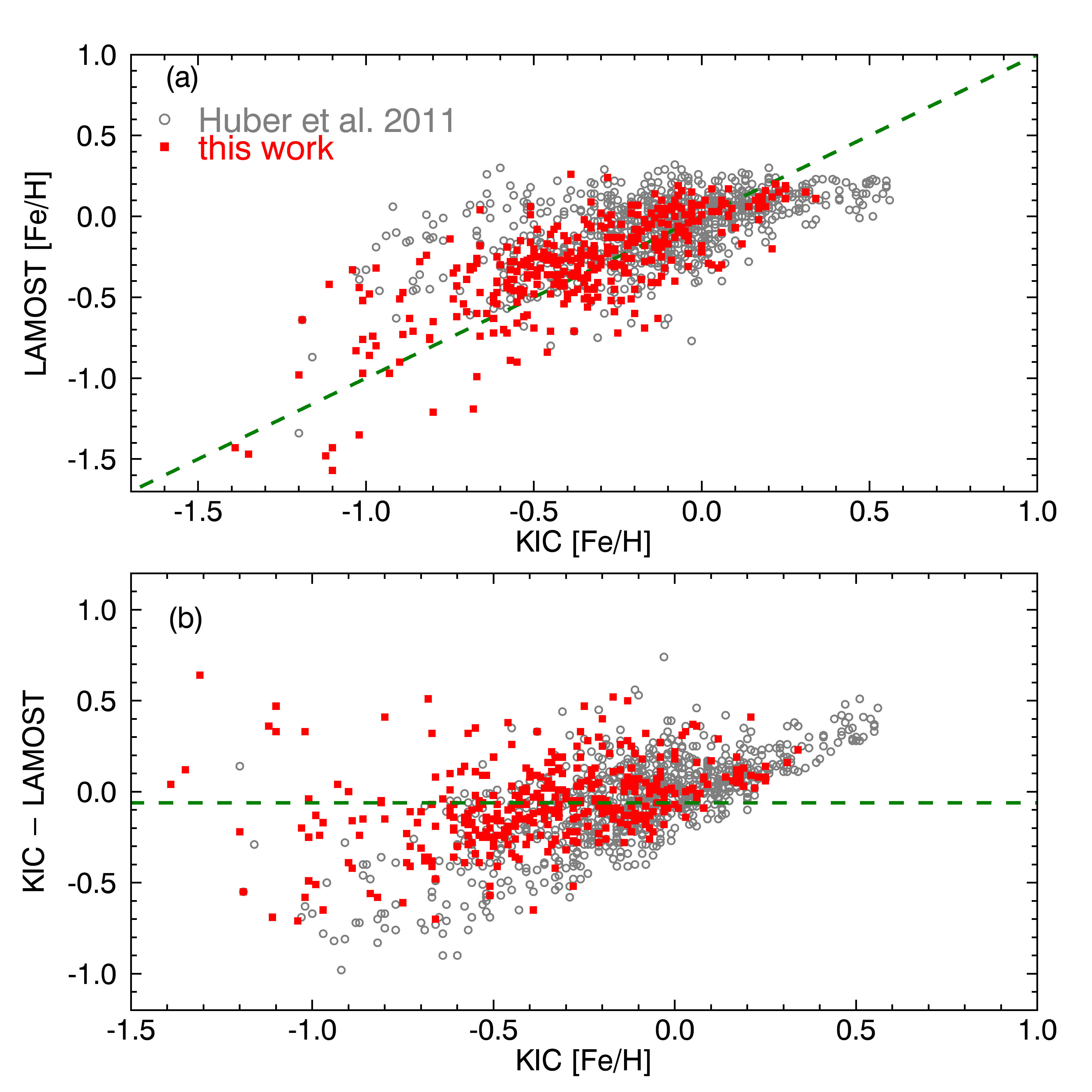}
\caption{Same with Figure \ref{teffcmp} but for metallicity.}
\label{metalcmp}
\end{center}
\end{figure}
Figure \ref{seislogg} compares the asteroseismic and LAMOST surface gravities against those from the KIC. Panel (a) clearly 
illustrates the misclassification of targets in our sample, where KIC surface gravities
 are greater than 3.4 dex and hence classified as subgiants while the seismic values are less than 3.5 dex and thus are classified as red giants (see Figures \ref{fnumaxdnu}, 
\ref{famplitude}, \ref{fhrdiagram} and \ref{fseishrdiagram}). 

In panel (b) we observe a similar distribution as shown in panel (a) but with larger scatter, which confirms the misclassification of our sample as subgiants by the KIC.
With respect to targets from \citet{Huber11}, the agreement in both panels is good overall but still has substantial spread, in particular for dwarfs in panel (b). The mean and scatter of the log~$g$ residuals are $-0.10\pm0.30$ dex~in the sense of seismic minus KIC and $-0.15\pm0.35$ dex~in the sense of LAMOST minus KIC. We are therefore able to conclude that the incorrect determination of KIC surface gravities for our sample is the main cause for the erroneous prediction of \numax. 
\subsection{Are the red giants in our sample physically different?}
Realizing that incorrect KIC surface gravities are responsible for the incorrect \numax~predictions makes us wonder whether some particular physical properties led to the 
misclassification of our sample in the KIC. 
To explore whether our sample and the \citet{Huber11} sample are physically different in metallicity, we illustrate the corresponding 
comparison in Figure \ref{metalcmp}. It shows that the two samples are similar in metallicity, spanning 
from -1.4 dex to 0.5 dex. Comparing with LAMOST, KIC metallicities of both samples were underestimated in the metal-poor regime but overestimated in the metal-rich region. 
In panel (b) we clearly observe this trend with a mean and standard deviation of $-0.061$ dex and 0.210 dex, respectively, in the sense of KIC minus LAMOST. This is consistent with 
the metallicity trend found by \citet{Dong14}. 
\begin{figure*}
\begin{center}
\includegraphics[scale=0.8]{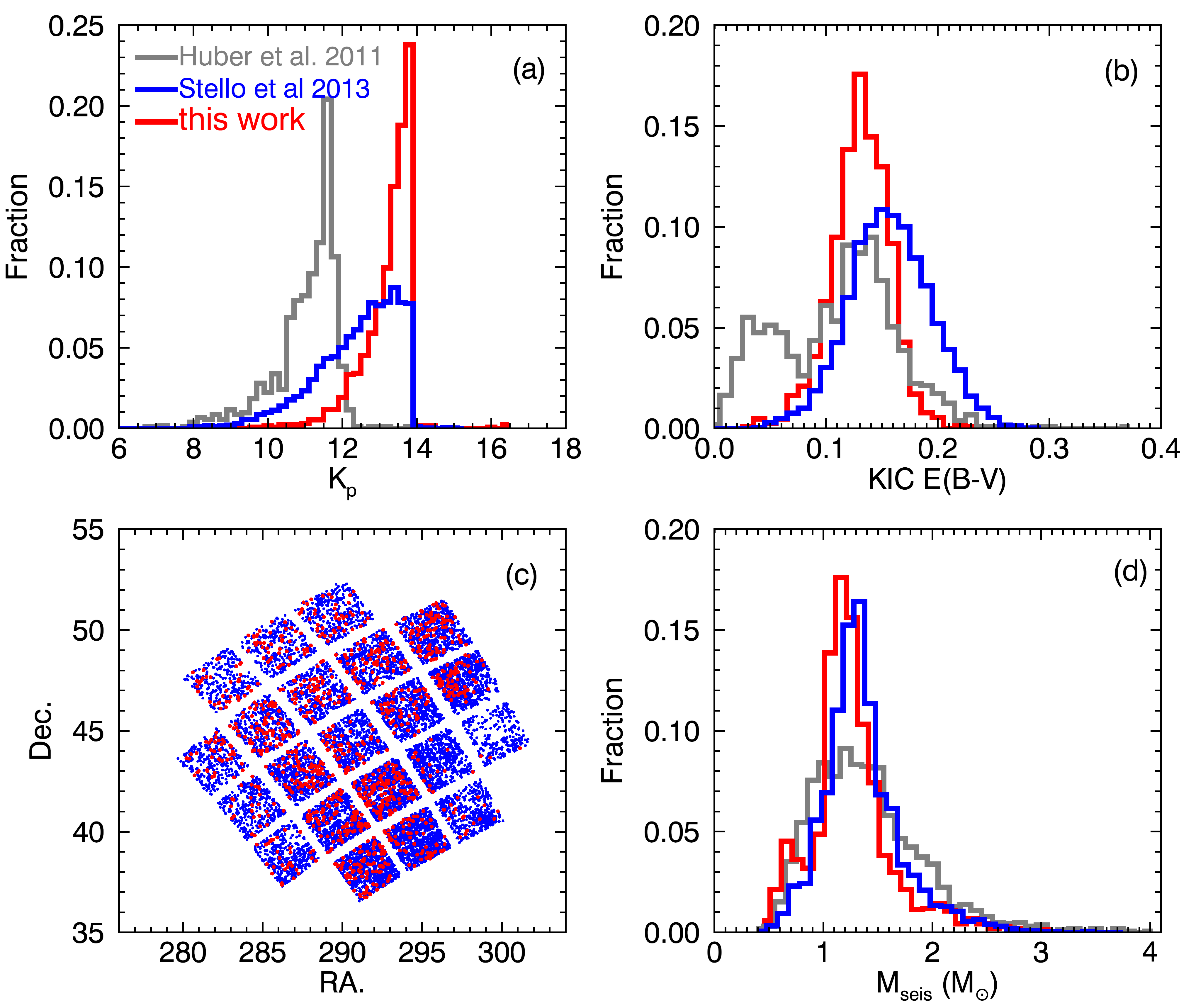}
\caption{Histograms for \kep~magnitude $K_p$ (panel (a)) and KIC reddening (panel (b)),  spatial distribution in \kep~field of view (panel (c)) and histogram of asteroseismic mass (panel (d)). 
The red data represents targets of our entire sample, the grey data indicates those from \citet{Huber11} while the blue data are targets from \citet{Stello13}.}
\label{fnormcmp}
\end{center}
\end{figure*}

We also checked if the lack of seven independent colors (derived from eight filters: KIC griz, D51 and 2MASS JHK) contributes to the incorrect KIC predicted \numax~values. 
However, only three stars do not have a full set of colors and therefore we excluded this effect as a dominant cause for the larger errors in log $g$.

In Figure \ref{fnormcmp}, we compare \kep~magnitudes, KIC reddening, asteroseismic masses and spatial distribution for our sample (red data) 
with the samples from \citet{Huber11} (grey data) and 
\citet{Stello13} (blue data). Note that we have restricted both comparison samples to seismic gravities in the same ranges as our sample (from 2.2 dex to 3.5 dex).

From Figure \ref{fnormcmp}a we can see, due to the preferential selection of bright stars by the Kepler Asteroseismic Science Consortium (KASC), our sample is systematically fainter by nearly 2 mag, 
compared to \citet{Huber11}. We note that \citet{Huber11} used data spanning from Q0 to 
Q6 for long-cadence light curves and Q0 to Q4 for short-cadence light curves, whereas we used data from the entire nominal \kep~Mission (Q0-Q17). Longer time series enhance the probability of detecting oscillations in fainter stars. Both our sample and the \citet{Stello13} sample show a sharp cut at $K_p = 14~$mag due to the \kep~selection function \citep{Batalha10}.
Panel (b) shows most targets in our sample have reddening around 0.13 mag, which is consistent with the sample of \citet{Huber11} but shifted by 0.03 mag compared to the \citet{Stello13} sample. 
Inspection of panel (b) also reveals there are a population of stars with lower reddening peaking at 0.035 mag. 
To understand if this population of stars reside in special location, 
we plotted the spatial distribution of targets for our sample (red dots) and \citet{Stello13} (blue dots) in the entire \kep~field of view. We see fewer targets located in the three modules at the bottom-right corner, which are closest to the Galactic plane and have less stars because of the \kep~targets selection function. 
Particularly, we checked that the population of stars with reddening less than 0.08 are almost uniformly distributed. Thus, we can conclude that sky position does not correlate with  
the incorrect KIC surface gravities.

It is tempting to speculate that the $\approx$ 0.03 mag systematic reddening shift between our sample and the \citet{Stello13} sample might be responsible for the wrong determination of surface gravities and hence 
lead to the incorrect KIC \numax~prediction. However, we should note that KIC effective temperatures show good agreement with the LAMOST values. Besides, it is also possible that the larger reddening values 
are a simple consequence of the fact that the \citet{Stello13} sample includes more evolved stars, which are on average more distant, keeping in mind the similar magnitude distributions for our sample and 
\citet{Stello13}.

In panel (d) we plot the histogram of seismic masses and find that stars for all three samples 
have asteroseismic masses peaking around 1.2 \msun, although stars from \citet{Huber11} have a slightly wider spread.

Therefore, we can conclude from Figure \ref{metalcmp} and Figure \ref{fnormcmp} that targets from this work and the previous studies are similar in terms of reddening, sky position, mass and metallicity. 
This indicates that our sample is not physically 
different compared to the \citet{Huber11} and \citet{Stello13} samples, and that the misclassification is most likely due to random errors in the KIC surface gravity values.

\section{Conclusions}
We selected 4758 stars with KIC predicted \numax~ranging from close to the Nyquist frequency (280 \muHz) to 700 \muHz. Based on the SYD pipeline 
applied to the long-cadence light curves observed by \kep, we detected unambiguous oscillations in 1523 red giants and reported the discovery of 626 new oscillating red giants.
 Our sample increases the known number of oscillating low-luminosity red giants by 26\%~(up to $\sim$~1900 stars) 
and includes 47 seismic targets residing in the 
super-Nyquist frequency region up to 387 \muHz. RGB stars account for approximately $70.5\%$ of the sample while RC1 stars take up $\approx$ 24.5$\%$ of sample. The remaining $\approx$ $5.0\%$ of stars 
can either be classified as RGB or RC2, but further discrimination needs more detailed analysis.

The significant difference between the observed and KIC predicted \numax~arises mostly from the incorrect determination of surface gravities. 
Comparison of asteroseismic surface gravities with those from the KIC clearly illustrates the misclassification 
of targets in our sample. Surface gravities returned from the KIC are greater than 3.4 dex for all stars while seismic values are less than 3.5 dex, thus classifying those stars as red giants.

We argue that the incorrect KIC surface gravities do not result from the physical properties such as reddening, spatial distribution, mass or metallicity between our sample
and those from \citet{Huber11} and \citet{Stello13}. We hence believe that our sample is not physically different compared to the one of \citet{Huber11}, but rather a result of misclassification due to large 
errors in the KIC.

The unambiguous oscillations are very valuable to understand stellar global properties and interior structure such as rotation and magnetic fields. The key synergies between asteroseismology and 
exoplanet science will also allow us to characterize planet candidates around those stars provided transit and oscillations can be detected simultaneously (Yu et al. in prep.). We are also in the process of compiling a full 
catalog of asteroseismic parameters for all oscillating red giants observed over the four-year \kep~Mission.
\section*{Acknowledgments}
The authors would like to thank the referee for a careful reading of our manuscript and many helpful comments.
We gratefully acknowledge the entire \kep\ team and everyone 
involved in the \kep\ mission for making this paper possible. 
Funding for the \kep\ Mission is provided by NASA's Science Mission Directorate. 
Some/all of the data presented in this paper were obtained from the Mikulski Archive for Space Telescopes (MAST). STScI is operated by 
the Association of Universities for Research in Astronomy, Inc., under NASA contract NAS5-26555. Support for MAST for non-HST data is 
provided by the NASA Office of Space Science via grant NNX09AF08G and by other grants and contracts. 
Guoshoujing Telescope (the Large Sky Area Multi-Object Fiber Spectroscopic Telescope, LAMOST) is a National Major Scientific 
Project which is built by the Chinese Academy of Sciences, funded by the National Development and Reform Commission, and operated 
and managed by the National Astronomical Observatories, Chinese Academy of Sciences.
D.H. acknowledges support by the Australian Research Council's Discovery Projects funding scheme (project number DE140101364) and 
support by the National Aeronautics and Space Administration under Grant NNX14AB92G issued through the Kepler Participating Scientist Program. T.D.L. acknowledges support by grant 11273007 from the National Natural Science Foundation of China.

\bibliographystyle{mn2e}
\bibliography{/Users/daniel/science/codes/latex/references}

\end{document}